\documentclass{article}

\usepackage{arxiv}
\usepackage[utf8]{inputenc} 
\usepackage[T1]{fontenc}    
\usepackage{hyperref}       
\usepackage{url}            
\usepackage{booktabs}       
\usepackage{amsmath,amsfonts,amssymb,amsthm,mathrsfs}
\usepackage{nicefrac}       
\usepackage{graphicx}
\usepackage{natbib}
\usepackage{doi}
\usepackage{comment}
\usepackage{empheq}
\usepackage[most]{tcolorbox}

\newtcbox{\mymath}[1][]{%
    nobeforeafter, math upper, tcbox raise base,
    enhanced, colframe=blue!30!black,
    colback=blue!10, boxrule=1pt,
    #1}

\newtheorem{thm}{Theorem}

\newcommand{\revision}[1]{{\color{black}#1}}

\title{\revision{Employing observability rank conditions for taking into account experimental information \textit{a priori}}}
\author{\href{https://orcid.org/0000-0001-7401-7380}{\includegraphics[scale=0.06]{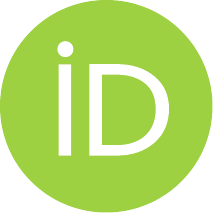}\hspace{1mm}Alejandro F.~Villaverde}\\
	CITMAga, 15782 Santiago de Compostela, Galicia, Spain\\
	Universidade de Vigo, Department of Systems Engineering and Control,
	36310 Vigo, Galicia, Spain \\
	\texttt{afvillaverde@uvigo.gal} \\
}

\date{}

\renewcommand{\headeright}{}
\renewcommand{\undertitle}{}
\renewcommand{\shorttitle}{}

\def\R{\ensuremath{\mathbb{R}}}

\def\nei{\ensuremath{\mathcal{N}}}
\def\O{\ensuremath{\mathcal{O}}}
\def\L{\ensuremath{\mathcal{L}}}

\hypersetup{
pdfkeywords={Identifiability, Observability, Dynamic modelling, Differential geometry},
}

\begin{document}
\maketitle

\begin{abstract}
	The concept of identifiability describes the possibility of inferring the parameters of a dynamic model by observing its output. It is common and useful to distinguish between structural and practical identifiability. The former property is fully determined by the model equations, while the latter is also influenced by the characteristics of the available experimental data. Structural identifiability can be determined by means of symbolic computations, which may be performed before collecting experimental data, and are hence sometimes called \textit{a priori} analyses. Practical identifiability is typically assessed numerically, with methods that require simulations -- and often also optimization -- and are applied \textit{a posteriori}. 
    An approach to study structural local identifiability is to consider it as a particular case of observability, which is the possibility of inferring the internal state of a system from its output. Thus, both properties can be analysed jointly, by building a generalized observability matrix and computing its rank.
    The aim of this paper is to investigate to which extent such observability-based methods can also inform about \revision{practical aspects related with the experimental setup, which are usually not approached in this way}. To this end, we explore a number of possible extensions of the rank tests, and discuss the purposes for which they can be informative as well as others for which they cannot.
 \end{abstract}


\section{Introduction}

\revision{Identifiability is a broad concept related to mathematical modelling that is used in several contexts, from statistics to econometrics to dynamical systems. It refers to the possibility of inferring the parameters of a model by observing its output. In the area of system identification it is common to use the term \textit{structural identifiability}. This concept} was introduced by \cite{bellman1970structural}, who used the word ``structure'' to refer to the equations of a dynamic model. This property determines whether it is possible ``to get insight into the internal structure of a system from input-output measurements''. When analysing structural identifiability, one imposes no restrictions on the quality and quantity of data that can be obtained from the experiments. Thus, if a certain state variable, or function, is defined as a model output, the implicit assumption is that it can be sampled at an arbitrarily high rate and with absolute precision. Since this is an ideal scenario, another concept -- practical identifiability -- is needed to account for the additional uncertainties caused by the limitations of real-world data, which are noisy and scarce to a varying degree \citep{lam2022practical}. 
Structural identifiability analyses provide a binary answer: a model is either structurally identifiable or unidentifiable. In contrast, practical identifiability analyses yield a numerical answer, typically in the form of a confidence interval in a parameter value. 
Overviews of practical and structural identifiability have been written by \cite{miao2011identifiability} and more recently by \cite{wieland2021structural}.

In this paper we equate ``structural'' with \textit{a priori}, and ``practical'' with \textit{a posteriori}. 
We note that other definitions, or distinctions between ``structural'' and ``practical'' approaches, can be found in the literature \citep{anstett2020priori}. Here we will not discuss in depth the subtleties related with these definitions, since they are not important for the purposes of the present work. Instead, we only need to clarify that the present investigation concerns the use of symbolic computation methods -- and specifically those based on a differential geometry approach -- to refine the identifiability results that can be obtained from the model equations only, by exploiting additional information about experimental constraints (without having to produce any experimental or synthetic data).

The observability of a system describes the possibility of inferring its internal state by observing its output. 
\revision{Here, internal state refers to the values at a given time of the set of time-varying variables given by the differential equations of the model.}
The concept was introduced by R. E. Kalman in the context of linear control systems. For nonlinear systems, \cite{hermann1977nonlinear} presented an observability test called observability rank condition; \cite{tunali1987new} applied a similar condition to the study of identifiability. It should be noted that, if a parameter in a model is considered as a constant state variable, (structural local) identifiability can be studied in the same way as observability. Both concepts were reviewed jointly in \citep{villaverde2019observability}.

In this paper we seek answers to the following question: what information, if any, can structural analyses -- and specifically, the differential geometry approach based on the rank test -- provide about practical identifiability? 
Our interest in the use of structural methods instead of numerical ones to analyse practical identifiability comes from three reasons.
First, while some numerical approaches may be faster than the structural ones, this is not always the case, especially since recent advances have greatly improved the efficiency of structural methods \citep{rey2023benchmarking}. Second, it is good modelling practice to analyse both structural and practical identifiability, one \textit{a priori} and the other \textit{a posteriori} \citep{villaverde2022protocol}, since these analyses provide different types of information that are valuable at different stages of the modelling process. And, since structural analyses may reveal inadequacies of the model structure that are difficult to learn otherwise, information about the exact way in which these limitations affect practical identifiability would be helpful.
Third, given that one should perform structural analyses anyway, it is seems wise to try and maximise the information that they yield. 

The remainder of this paper is structured as follows: first, we provide some definitions and known results in Section \ref{sec:background}. Then, we explore a number of possible extensions to classic differential geometry methods in Section \ref{sec:bridging}. We finish the paper with some concluding remarks, including ideas for future work, in Section \ref{sec:conclusions}.

\section{Background}\label{sec:background}

\subsection{Structural identifiability and observability}\label{sec:sio}

The methods described in this paper can be applied to dynamic models described by ordinary differential equations of the following form:

\begin{equation}\label{mod}
   M:\left\{\begin{aligned}
    \dot{x}(t) & = & f\left(u(t),x(t),\theta\right), \\
    y(t) & = & g\left(u(t),x(t),\theta\right),
\end{aligned}\right. 
\end{equation}
 where $f$ and $g$ are possibly nonlinear analytic functions; $x(t)\in\R^{n_x}$ is the vector of state variables; $u(t)\in\R^{n_u}$ is the vector of (infinitely differentiable) inputs; $y(t)\in\R^{n_y}$ is the output vector; and $\theta\in\R^{n_\theta}$, the parameter vector. The input and output vectors are assumed to be perfectly known; the parameters and state variables are in principle unknown. Whenever convenient, we will use a simplified notation and omit the dependency of $u$, $x,$ and $y$ on $t.$

We say that a parameter $\theta_i\in\theta$ of $M$ (\ref{mod}) is \textit{structurally locally identifiable} (SLI) if\revision{, for almost any vector $\theta^*\in\R^{n_\theta}$,} there is a neighbourhood $\nei(\theta^*)$ such that, for any $\hat{\theta}\in\nei(\theta^*)$, $y(t,\theta^*)=y(t,\hat{\theta}) \iff \theta^*_i=\hat{\theta}_i$ \revision{\citep{walter1997identification,chis2011structural,distefano2015dynamic}}. 
If this relationship does not hold in any $\nei(\theta^*)$, $\theta_i$ is structurally unidentifiable (SU). \revision{The phrase ``almost any vector'' means that the only possible exceptions belong to a set of measure zero.}

The above definition entails that, if a parameter is SLI, its value can be determined from knowledge of the output $y(t)$ and input $u(t)$ vectors over time.
Structural local identifiability is similar to another property, observability \revision{\citep{chatzis2015observability}}. Conceptually, a state variable $x_i(\tau)$ is \textit{observable} if it can be determined from the output $y(t)$ and any known inputs $u(t)$ of the model in the interval $t_0 \leq \tau \leq t \leq t_f$, for a finite $t_f$. Otherwise, it is unobservable. 
A model is called SLI if all its parameters are SLI, otherwise it is said to be SU.
Likewise, a model is called observable if all its states are observable, and unobservable if at least one of them is unobservable \revision{\citep{wieland2021structural}}.

It should be noted that the properties defined above are structural but \textit{local}, i.e. they tell us whether a parameter or a state variable can be uniquely estimated in a neighbourhood of (almost) any particular value, but they do not guarantee uniqueness in the whole of $\R^{n_x}$ or $\R^{n_\theta}.$ 
In contrast, a parameter is said to be structurally globally identifiable (SGI) if, for any $\hat{\theta}\in\nei(\theta^*)$ and almost any vector $\theta^*\in\R^{n_\theta}$, $y(t,\theta^*)=y(t,\hat{\theta}) \iff \theta^*_i=\hat{\theta}_i$, i.e. the neighbourhood $\nei(\theta^*)$ is the whole parameter space. A SGI parameter is SLI; a SLI parameter may be SGI \revision{\citep{walter1997identification}}. 

Structural identifiability can be analysed from an algebraic point of view (i.e. looking at the relations among variables in the equations) or from a geometric point of view (i.e., broadly speaking, looking at the shape of the solutions). 
Each of these approaches has its advantages and limitations. Differential algebra methods \citep{ljung1994global,saccomani2001new} can characterise local and global identifiability, but they are only applicable to rational models. On the other had, a differential geometry approach based on Lie derivatives can analyse models with non-rational terms, but it can only inform about local identifiability and observability. 

We may wonder, then, how important this distinction between local and global identifiability is in real applications. In other words: if a differential geometry technique reports that a parameter is SLI, how likely is it that it is also SGI? And, if it is SLI but not SGI, how likely is it to obtain a wrong estimate? A partial answer to these questions was given in \citep{barreiro2023origins}, where an empirical study on over a hundred mathematical models of biological systems with a total of more than seven hundred parameters was conducted. The results showed that approximately 5\% of the parameters were locally, but not globally identifiable, a property for which the acronym SLING was coined. Furthermore, only for 3\% of the parameters it was possible to obtain a wrong estimate, since in the remaining 2\% the alternative local solutions did not have biological meaning (e.g. their values were either negative, or far from the feasible range). 

In the remainder of this paper we will use an approach for analysing structural local identifiability and observability that is based on differential geometric concepts, namely Lie derivatives. The next subsection describes this approach.

\subsection{A rank test for assessing observability}

We start by studying the observability of a nonlinear model of the form:
\begin{equation}\label{mod2}
   M:\left\{\begin{aligned}
    \dot{x} & = & f\left(x,u\right), \\
    y & = & g\left(x,u\right).
\end{aligned}\right. 
\end{equation}
Such a model is similar to the one given by \eqref{mod}, with the difference that it does not contain parameters (we will include them in the next subsection). 
\revision{There are several ways of analysing its observability. The core intuition that underlies some of them can be summarized as follows: first, define an object that contains the information that can be directly obtained from measuring the output, i.e., the output itself $y$ and its derivatives, $\dot{y},\ddot{y},...$.
Then, establish a link between this mapping and the internal state of the system.
Lastly, determine whether this mapping is unique, which would mean that the internal state can be derived from the output information. 

A way of performing this test is to use the Taylor series expansion of the output function around the initial condition $x(t_0)$, and then check whether this expansion is unique. This idea was proposed by \cite{pohjanpalo1978system} for the purpose of structural identifiability -- not observability -- analysis. To perform this analysis, one must first obtain a set of algebraic equations on the parameters. If their solution is unique, then the parameters are SGI.
A conceptually similar approach is the Generating Series introduced by \cite{walter1982global}. Here, the output is expanded in series with respect to time and the inputs, whose coefficients are the output functions and their \textit{Lie derivatives}.
The Lie derivative $\L_fg(x,u)$ of the output function $g$ along the vector field $f$ is defined as:
\[
\L_fg(x,u)=\frac{\partial g(x,u)}{\partial x}f(x,u)+\sum_{j=0}^{j=\infty}\frac{\partial g(x,u)}{\partial u^{(j)}}u^{(j+1)}.
\]
The Lie derivative quantifies the rate of change of a vector field along the flow described by another vector field. In this case, it describes the change of the model output function $g$ along the dynamics of the system $f$, which is equal to the derivative of the output, $\dot{y}.$
Subsequent derivatives of the output, $\ddot{y},..., y^{(n_x-1)},$ are calculated as higher order Lie derivatives, which can be computed recursively from the lower order ones as follows \citep{karlsson2012efficient}:
\begin{equation}\label{eq_Lie}
\L_f^ig(x,u)=\frac{\partial \revision{\L_f^{i-1}}g(x,u)}{\partial x}f(x,u)+\sum_{j=0}^{j=\infty}\frac{\partial \revision{\L_f^{i-1}}g(x,u)}{\partial u^{(j)}}u^{(j+1)}
\end{equation}
In the generating series approach, one builds an \textit{exhaustive summary} that consists of the coefficients of the output and their successive Lie derivatives up to some order, evaluated at the initial condition $x(t_0)$. If the exhaustive summary is unique, then the parameters are SGI.

Here we adopt an approach that has similarities to the Taylor and Generating Series methods outlined above. It is sometimes called the \textit{geometric} approach, and it follows the line of work initiated in the 1970s by a number of researchers, including \cite{griffith1971observability,kou1973observability,hermann1977nonlinear}. 
It begins by defining an ``observability mapping'' that consists of the output and its time derivatives, and it is obtained by calculating $n_x-1$ Lie derivatives of $g(x,u)$. Then, the relation between the observability mapping and the internal state of the system is obtained by computing an ``observability matrix'', $\O(x,u)$. This matrix consists of the Jacobian of the observability mapping with respect to the state variables, i.e.:}
 \begin{equation}\label{obsmat2}
   \O(x,u)=\begin{pmatrix}
    \frac{\partial}{\partial x}y\;\; \\
    \frac{\partial}{\partial x}\dot{y}\;\;\\
    \vdots\qquad\\
    \frac{\partial}{\partial x}y^{(n_x-1)}
    \end{pmatrix}
    =\begin{pmatrix}
    \frac{\partial}{\partial x}g(x,u)\quad \;\;\\
    \frac{\partial}{\partial x}\left(\L_fg(x,u)\right)\;\;\\
    \vdots\qquad\quad\\
    \frac{\partial}{\partial x} \left(\L_f^{n_x-1}g(x,u)\right)
\end{pmatrix}.
\end{equation}

Then, observability is assessed by calculating the rank of $\O(x,u)$:
\begin{thm}[Observability Rank Condition, ORC \citep{hermann1977nonlinear}]
    A model $M$ given by \eqref{mod} is observable around a generic point $x_0$ if and only if $\text{rank}(\O(x_0,u))=n_x$.
\end{thm}

\revision{A note about the calculation of the Lie derivatives is in order. Note that the expressions in \eqref{eq_Lie}} contain infinite summations. However, since $\dot u$ does not appear in the model equations \eqref{mod}, the summations can be truncated so that \citep{villaverde2019full}:
\begin{align}\label{eq:Lie1}
\L_fg(x,u) & =\frac{\partial g(x,u)}{\partial x}f(x,u),\\ \label{eq:Lie2}
\L_f^ig(x,u) & =\frac{\partial L_f^{i-1} g(x,u)}{\partial x}f(x,u)+\sum_{j=0}^{j=i-1}\frac{\partial g(x,u)}{\partial u^{(j)}}u^{(j+1)}.
\end{align}

\subsection{Analysing structural identifiability as observability}\label{sec:sio_an}

Let us now consider a model with parameters such as \eqref{mod}.
According to the definitions provided in Section \ref{sec:sio}, a state variable is observable if its initial condition can be inferred by measuring the model output at later times. Since a parameter can be considered as a state variable with a constant value \citep{tunali1987new}, its structural local identifiability can be analysed using the observability method that we have just described. 
To this end we augment the state vector as $\tilde{x}=\left[x;\theta\right],$ which now has dimension $n_{\tilde{x}}=n_x+n_\theta,$ and the corresponding vector of states derivatives is $\dot{\tilde{x}}=\left[f(\tilde{x},u);\;\mathbf{0}\right]$ (where the use bold font to indicate that the zero is a vector of size $n_\theta$). 
For a new model with these changes in the variables and equations -- which is equivalent to the original one -- we can build an observability-identifiability matrix, $\O_I(\tilde{x},u)$, in the same way as the $\O(x,u)$ of \eqref{obsmat2}, that is:
 \begin{equation}\label{obsidmat}
   \O_I(\tilde{x},u)=\begin{pmatrix}
    \frac{\partial}{\partial \tilde{x}}y\;\; \\
    \frac{\partial}{\partial \tilde{x}}\dot{y}\;\;\\
    \vdots\qquad\\
    \frac{\partial}{\partial \tilde{x}}y^{(n_x+n_\theta-1)}
\end{pmatrix}=\begin{pmatrix}
    \frac{\partial}{\partial \tilde{x}}g(\tilde{x},u)\quad \;\;\\
    \frac{\partial}{\partial \tilde{x}}\left(\L_fg(\tilde{x},u)\right)\;\;\\
    \vdots\qquad\quad\\
    \frac{\partial}{\partial \tilde{x}} \left(\L_f^{n_x+n_\theta-1}g(\tilde{x},u)\right)
\end{pmatrix}.
\end{equation}

Then, if system $M$ given by (\ref{mod}) satisfies $\text{rank}(\O_I(\tilde{x}_0,u))=n_{\tilde{x}}=n_x+n_\theta$, with $\tilde{x}_0$ a point in the augmented state space, the model is observable and identifiable around $\tilde{x}_0$. We refer to this property with the acronym SIO (structural identifiability and observability).

If $\text{rank}(\O_I(\tilde{x}_0,u))<n_{\tilde{x}}=n_x+n_\theta,$ the model contains at least one unobservable state variable (which may be in fact an unidentifiable parameter considered as a state variable). To determine which ones, we note that the $i^{\text{th}}$ column of $\O_I(\tilde{x}_0,u)$ is obtained by taking the partial derivative of the output (and its time derivatives) with respect to the $i^{\text{th}}$ variable: 

\begin{empheq}[box=\mymath]{equation}\label{obsumat}
%
\mathcal{O}_I(\tilde x,u)=\begin{pmatrix}
    \frac{\partial \mathcal{L}_{\hat{f}}^0g}{\partial \tilde x_1} & \frac{\partial \mathcal{L}_{\hat{f}}^0g}{\partial \tilde x_2} & \dots & \frac{\partial \mathcal{L}_{\hat{f}}^0g}{\partial \tilde x_{n_x+n_\theta}} \\
    \frac{\partial \mathcal{L}_{\hat{f}}g}{\partial \tilde x_1} & \frac{\partial \mathcal{L}_{\hat{f}}g}{\partial \tilde x_2} & \dots & \frac{\partial \mathcal{L}_{\hat{f}}g}{\partial \tilde x_{n_x+n_\theta}} \\
    \vdots & \vdots & \ddots & \vdots  \\
    \frac{\partial \mathcal{L}_{\hat{f}}^{n_x+n_\theta-1}g}{\partial \tilde x_1} &  \frac{\partial \mathcal{L}_{\hat{f}}^{n_x+n_\theta-1}g}{\partial \tilde x_{2}} & \dots & \frac{\partial \mathcal{L}_{\hat{f}}^{n_x+n_\theta-1}g}{\partial \tilde x_{n_x+n_\theta}}
\end{pmatrix}.
\end{empheq}

Hence, the SIO of an individual variable can be determined as follows: remove column $i$ from $\mathcal{O}_I$ and calculate the rank of the resulting matrix, $\O^{-i}_I$. If $rank(\O^{-i}_I) < rank(\O_I),$ the $i^{th}$ variable is observable (i.e., if it is a parameter, it is SLI). If $rank(\O^{-i}_I) = rank(\O_I),$ it is unobservable (SU).

Lastly, it should be noted that it is not always necessary to calculate $n_{\tilde{x}}-1$ Lie derivatives, since a matrix $\O_I$ built with a smaller number may still have full rank. The minimum number of Lie derivatives for which this may happen, $n_d$, is the number for which the matrix has at least as many rows as columns, which is given by $n_d = \left\lceil \frac{n_x+n_{\theta}-n_y}{n_y} \right\rceil$. Therefore, it is possible to use a step-wise procedure for for building $\O_I,$ computing the rank after $n_d$ derivatives and adding more derivatives it if it is not full, which may produce computational savings. This idea is implemented e.g. in the STRIKE-GOLDD toolbox \cite{diaz2022strike}.

\subsection{Practical Identifiability}

While a model's SIO is fully determined by its equations, practical identifiability (PI) is also affected by experimental conditions. Structural identifiability is a necessary but not sufficient condition for PI. The goal of PI analysis is to quantify how much parameter estimates vary due to limitations in data quantity and quality. Following \cite{lam2022practical}, we may classify PI methods in four main groups: sensitivity-based, profile likelihood, Monte Carlo, and Bayesian. 

Briefly, sensitivity-based methods typically obtain bounds for the parameter estimates from the Fisher Information Matrix (FIM), which is calculated from a sensitivity matrix that measures the dependence of the output on parameter values at different time points \citep{rodriguez2006novel}. Information about parameter correlations can also be obtained in this way. 
Monte Carlo approaches are based on solving the inverse problem many times, using different noise realisations from repeated simulations \citep{balsa2010iterative}. 
The profile likelihood approach generates cost curves for each parameter by fixing it to some value around its optimum and re-optimising the remaining parameters, and repeating this process for a range of values \citep{raue2009structural,simpson2023profile}.
Lastly, Bayesian methods provide probability distributions and incorporate prior belief about parameter values \citep{hines2014determination}.

As in the case of structural analyses, which can deal jointly with identifiability and observability (SIO), it is possible to use the methods of practical identifiability analysis for the study of practical observability; we may then use the acronym PIO.

The methods described above require multiple numerical simulations of the model in order to generate artificial data, and are therefore called \textit{a posteriori} approaches. In the next section we explore how the \textit{a priori} approach described in Section \ref{sec:sio_an} can (or cannot) inform about PIO without performing simulations.

\section{Modifying the rank test to take into account \revision{experimental information}}\label{sec:bridging} 

In this section we investigate in which ways we can use structural analyses -- and more specifically the observability-based approach described in Sections \ref{sec:sio}--\ref{sec:sio_an} -- to obtain information about identifiability \revision{for specific experimental configurations}. Thus, our investigation will be centred around one object, the observability-identifiability matrix $\O_I$ defined in \eqref{obsumat}, and we will discuss a number of directions in which to extend it or modify it so as to take into account `practical' aspects of the estimation process.

\subsection{Multiple experiments}\label{sec:multi}

The rank condition informs about the limits of observability and identifiability when data is collected from a  single experiment.
For linear models with one output, single-experiment and multiple-experiment identifiability coincide. However, for nonlinear models with multiple outputs, increasing the number of experiments could make more parameters or functions of parameters identifiable.

\cite{ovchinnikov2022multi} presented an algorithm that computes 
(i) the smallest number of experiments to maximise local identifiability, and (ii) the number of experiments to maximise global identifiability so that this number exceeds the minimal such number by at most one. This algorithm \revision{combines a differential algebra approach with observability rank checks.}  

With a differential geometry approach, information about multiple experiments can be incorporated into the ORC by using a modification of the $\O_I$ matrix of \eqref{obsumat} \citep{ligon2018genssi}. The idea is to create as many replicates of the output vector $y$ as experiments, as well as replicates of $x$ (if the initial conditions change among experiments) and $u$ (if the inputs change among experiments). Parameters $\theta$ are usually common across experiments. Depending on the characteristics of the model and the experiments, the ratio between outputs and variables can increase with additional experiments, possibly improving identifiability and/or observability.

Furthermore, having multiple experiments may also decrease the number of derivatives needed to build $\O_I,$ thereby reducing the computational cost of the analysis: as we have seen, the minimum number of Lie derivatives for which $\O_I$ may have full rank is $\left\lceil \frac{n_x+n_{\theta}-n_y}{n_y} \right\rceil;$ with $n_e$ experiments the number decreases to $\left\lceil \frac{n_e\cdot n_x+n_{\theta}-n_e\cdot n_y}{n_e\cdot n_y} \right\rceil.$

At least two software tools include this feature: GenSSI  \citep{ligon2018genssi} and STRIKE-GOLDD \cite{martinez2020nonlinear}; both of them are implemented in Matlab.

\subsection{Input characteristics}\label{sec:inputs}

Some entries of the $\O_I$ matrix may contain functions of the inputs and their derivatives, which appear in the formulas of the Lie derivatives \eqref{eq:Lie2}. When analysing identifiability, it is common to assume that the inputs are `sufficiently exciting', i.e. that they will induce dynamics leading to highly informative output data. However, in practice the type of inputs that can be applied is typically limited. For example, in many applications only constant inputs are possible, while in others at most ramp inputs may be applied. To a certain extent, these limitations can be taken into account by limiting the number of input derivatives that are nonzero \cite{villaverde2018input}. If this number is not infinite, the underlying assumption is that the input can be expressed as a polynomial function of a certain degree. Of course, it is also possible to perform the analyses for a particular input function.  

This idea can be combined with the multi-experiment setup, for example to study the effect of replacing a single experiment (with a time-varying input) with multiple experiments (with constant inputs).

\subsection*{* Piecewise constant inputs: multi-experiment or time-varying signals?}

A question related to Sections \ref{sec:multi} and \ref{sec:inputs}
concerns the implementation of inputs in real applications, and how to modify the analyses to approximate them. For example, instead of applying an ideal ramp, due to experimental constraints we may be forced to apply instead a piecewise constant input. How should we treat this scenario in our analyses? One the one hand, an input consisting of a series of constant values resembles a set of multiple experiments with constant inputs. On the other hand, as the number of steps increases the input becomes closer to an actual ramp. 
\revision{
\cite{wang1989two} showed that the observation spaces defined by means of piecewise constant inputs are equivalent to those defined by differentiable inputs. This result supports the idea of adopting the latter of the two aforementioned views--albeit with the caveats that usually apply to the extrapolation of results from ideal experiments to real ones.
}
The following example illustrates this case.

Consider the following linear, two-compartment model:
\begin{equation}\label{eq:comp}
\mathcal{M}_{3.2}:
\left\{%
\begin{array}{ll}
\dot{x_1} &= -(k_{1e} + k_{12})\cdot x_1 + k_{21}\cdot x_2 + b\cdot u,\\
\dot{x_2} &= k_{12}\cdot x_1 - k_{21}\cdot x_2,\\
y         &=  x_1
\end{array}
\right.
\end{equation}
The rank test reports that this model is fully identifiable and observable with generic inputs, and as a matter of fact for any polynomial inputs of degree at least one (i.e. a ramp input), i.e. as long as $\dot u \neq 0.$ However, all its parameters are unidentifiable if $\dot u = 0,$ both for a single experiment \citep{villaverde2018input} and in a multi-experiment setup with different constant inputs. 

\begin{figure}[htb]
    \centering
    \includegraphics[width=1.0\textwidth]{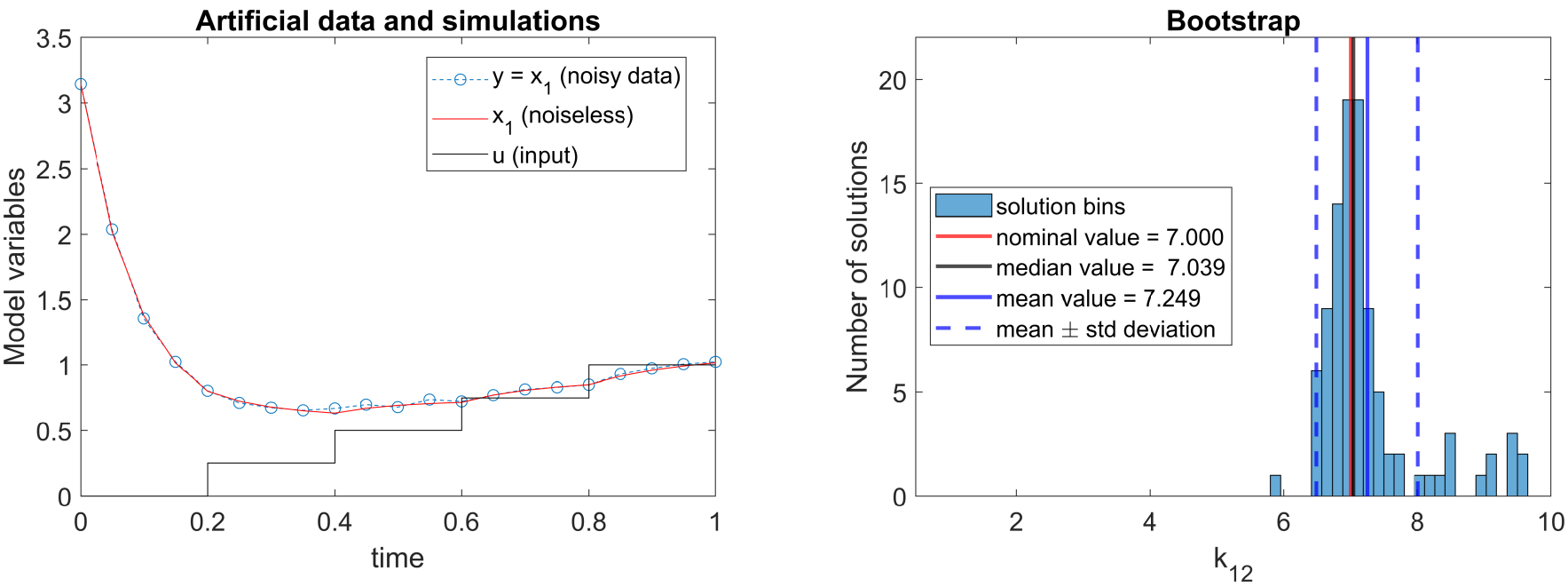}
    \caption{Model $\mathcal{M}_{3.2}$. The panel on the left shows simulations of the model input, $u(t),$ and the output $y(t)=x_1(t);$ the blue circles represent the artificial noiseless data used for parameter estimation. \revision{The panel on the right shows the resulting bootstrap of parameter $k_{12}$.}}
    \label{fig:c2m}
\end{figure}

Let us now perform a numerical experiment: we simulate the model to generate artificial data, applying as input a piecewise constant signal. Since our focus is on the role played by the input, \revision{we add a small amount of noise to the output, in order to yield a realistic experiment while avoiding confounding effects. We generate 100 realizations of Gaussian noise with zero mean and $\sigma = 0.1$.   
Next, we estimate the parameters by fitting the model to each realization of the data, using a hybrid metaheuristic optimization algorithm \citep{egea2014meigo}. Lastly, we use a bootstrap procedure to quantify their uncertainty \citep{balsa2010iterative}. 
The left panel of Figure \ref{fig:c2m} shows one realization of the pseudo-experimental data, along with the model states, and the input. 
The right hand panel of Figure \ref{fig:c2m} shows the optimization results for one of the parameters, $k_{12}$.  
It can be seen that, while there is some uncertainty associated with the estimate, both the mean and median estimations are very close to the nominal value ($k_{12}=7$), and the confidence intervals are reasonably narrow. It is clear from these results that $k_{12}$ is identifiable.}
Hence, performing the observability rank test by approximating a piecewise constant input as a ramp yields correct SIA results (at least in this case study), while the approximation as a multi-experiment with constant inputs yields a wrong result. Intuitively, this can be understood as a failure of the multi-experiment setup to capture the continuity between experiments; a set of independent experiments does not provide all the information.

\revision{
\subsection*{* Unknown inputs}
A related issue of practical importance concerns unknown inputs. So far we have assumed that the inputs $u(t)$ are perfectly known, but in reality we may face challenges such as non-negligible dynamic disturbances, unknown parameters that are time-varying, or external inputs that are known only to a certain extent. These problems can be seen as different ways of approaching the existence of unknown inputs. They do not only make estimation more difficult, but also complicate the observability analyses. 

In recent years, a number of results have been published that extend the ORC to enable the inclusion of unknown inputs \citep{martinelli2022extension,maes2019observability}. These works consider the unknown inputs and their derivatives, up to a given order, as additional state variables. The idea of limiting the number of nonzero input derivatives is similar to the assumption discussed for known inputs, although the computational cost is higher in the unknown input case. An efficient algorithm to check observability of models with unknown inputs was presented by \cite{shi2022efficient}.
}

\subsection{Initial conditions}

The definitions of structural local identifiability and observability are valid `almost everywhere', that is, with the possible exception of a set of measure zero. Thus, a model that is SLI may become unidentifiable for specific initial conditions. 
\revision{
The simplest example of this phenomenon is the model given by $\dot x = p\cdot x,$ $y=x.$
It is obvious that $x$ is observable and $p$ is SLI, except if $x(0)=0,$ in which case the system remains at the initial state and it is not possible to infer $p.$
Let us check the ORC for this model. Augmenting the state as explained in Section \ref{sec:sio_an}, we obtain
\begin{equation}\label{33a}
\mathcal{M}_{3.3.A}:
    \left\{%
    \begin{array}{ll}
    \dot{x}_1 = x_1\cdot x_2,\\
    \dot{x}_2 = 0, \\
    y_1 = x_1.
    \end{array}%
    \right.
\end{equation}
Since the model has two state variables, one Lie derivative suffices to obtain the observability-identifiability matrix as follows:
\begin{equation}
    \O_I=
    \begin{pmatrix}
        1   & 0 \\
        x_2 & x_1
    \end{pmatrix}
\end{equation}
The rank of this matrix is two, except for $x_1=0.$ Thus, the ORC result is in agreement with the loss of identifiability from the specific initial condition.
}
This issue is not rare, and it is typically linked to a lack of accessibility from the problematic initial conditions. (Accessibility, or reachability, is the possibility of moving from an initial point in parameter space to a set with a non-empty interior.) \cite{saccomani2003parameter} studied it using a differential algebra approach.
The following model \citep{august2009new} \revision{provides a more realistic} example:
\begin{equation}\label{michaelis}
\mathcal{M}_{3.3.B}:
    \left\{%
    \begin{array}{ll}
    \dot{x}_1 =-x_1\cdot x_2 + p_2\cdot(10-x_2),\\
    \dot{x}_2 =-x_1\cdot x_2 + (p_2 + p_3)\cdot(10 - x_2),\\
    \dot{x}_3 =-p_1\cdot x_3 + p_3\cdot(10 - x_2), \\
    y_1=x_1, y_2=x_3.
    \end{array}%
    \right.
\end{equation}
This model has six variables and $\text{rank}(\O_I)=6$, so it is SLI and observable. However, if $\{x_1(0)=0,\ x_2(0)=10\}$, it is clear that both states are always zero, and the only parameter that appears in the equations is $p_1,$ so the others become unidentifiable.
 
\revision{An important warning should be made here. In the example \eqref{33a}, we computed the ORC from a given initial condition simply by replacing the symbolic state variables in the matrix, $x_i,$ with specific numerical values. In this way, it is possible to test whether the model remains SLI from a given state. 
}
However, in doing so \textit{the stop criterion of the ORC is lost}: it may no longer be sufficient to build $\O_I$ with $n_x+n_{\theta}-1$ Lie derivatives, as in \eqref{obsumat}. The following case study\footnote{We thank Gleb Pogudin for noting this issue and providing the counter-example.} illustrates this fact. 

\begin{equation}\label{gleb}
\mathcal{M}_{3.3.C}:
    \left\{%
    \begin{array}{ll}
    \dot{x} =1+\theta \cdot x^2,\\
    x(0)=0,\\
    y=x.
    \end{array}%
    \right.
\end{equation}
This is a very simple model with just one state and one parameter; building its $\O_I$ with one Lie derivative (which is all we need because $n_x+n_{\theta}-1=1$) yields:

\begin{equation}
    \O_I=
    \begin{pmatrix}
        1 & 0 \\
        2\cdot\theta\cdot x & x^2
    \end{pmatrix}
\end{equation}
This matrix has rank = 2 if $x\neq 0,$ and rank = 1 if $x=0$. Therefore, $\theta$ appears to be unidentifiable from $x(0)=0.$ However, note that the third derivative of the output is
$$ \dddot y = \ddot{\left(\dot{x\ \ \ }\right)} = 2\cdot\theta\cdot x\cdot \ddot x + 2\cdot\theta\cdot\dot x \cdot \dot x,  $$
which, evaluated at $t=0,$ yields $y(0)=2\cdot\theta,$ so $\theta$ is clearly identifiable.

In fact, if we calculate the observability-identifiability matrix with three derivatives instead of one, we obtain
\begin{equation}
    \O'_I(x=0)=
    \begin{pmatrix}
        1 & 0 \\
        0 & 0 \\
        2\cdot\theta & 0 \\
        0 & 2
    \end{pmatrix},
\end{equation}
which has rank = 2, so a rank test performed with this matrix would yield the correct result, i.e. that $\theta$ is SLI. However, this would require knowing how many Lie derivatives should be calculated. In absence of that criterion, when generic state variables are replaced with specific values, the rank test gives a sufficient but not necessary condition for structural local identifiability and observability.

\subsection{Output derivatives}

The analysis of structural identifiability and observability assumes perfect knowledge of the output, and therefore of its derivatives. In real applications it is typically possible to measure only $y(t),$ and sometimes also $\dot y(t);$ however, higher order derivatives, $\ddot y(t), \dddot y(t), \ldots$ would need to be estimated. Unfortunately, experimental data in biology is often sparse and noisy, so accurate estimation of many derivatives is not possible.
Since the observability rank tests require building the $\O_I$ from (possibly many) output derivatives, one might wonder to what extent its results are a faithful reflection of the identifiability of the model in practice. 
To address this issue, \cite{thompson2022new} introduced the term ($v_1,\ldots,v_m$)-identifiability to refer to an analysis of identifiability that assumes that only the first $v_i$ derivatives of the $i^{th}$ output are available.
To assess ($v_1,\ldots,v_m$)-identifiability, they proposed to build the $\O_I$ matrix \eqref{obsumat} with only $v_i$ derivatives per output measure. Their analyses showed that the number of `v-identifiable' parameters in a model can increase drastically with $v_i$.
As an example, consider the following model (which is Example 2.1 in their paper):
\begin{equation}\label{2.1}
\mathcal{M}_{3.4}:
    \left\{%
    \begin{array}{ll}
    \dot x_1 &= \theta_1\cdot x_2,\\
    \dot x_2 &= -\theta_1\cdot x_1,\\
    \dot x_3 &= \theta_1\cdot\theta_2,\\
    y &= x_1+x_3.
    \end{array}%
    \right.
\end{equation}

In order to estimate $\theta_1,$ one can use the following relation, which is readily obtained by differentiating the output: 
\begin{equation}\label{eq21}
    \theta_1 = \pm \sqrt{\frac{-y^{(4)}(t_0)}{\ddot y(t_0)}} 
\end{equation}
Thus, we can infer $\theta_1$ from knowledge of the output and of its fourth derivative at a given time $t_0.$ Since in practice the measurement of $y(t)$ may be too noisy to allow for an accurate estimation of $y^{(4)}(t),$ this would seem to imply that $\theta_1$ would not be identifiable.
Likewise, in order to analyse its identifiability with the ORC, we need to build $\O_I$ with four Lie derivatives, and the rank test says that $\theta_1$ is identifiable. In contrast, if we limit the number of derivatives to three, the ORC says that $\theta_1$ is unidentifiable.  

However, we argue that the limited availability of derivatives does not impose such a strong limitation as claimed in \citep{thompson2022new}. First, as we saw in Section \ref{sec:multi}, with multiple experiments, less derivatives may be necessary. As a matter of fact, model $\mathcal{M}_{3.4}$ above requires building $\O_I$ with four derivatives to achieve full rank from a single experiment, but only three derivatives with two or more experiments.
Furthermore, while equation \eqref{eq21} relates $\theta_1$ to $y^{(4)}$ through a single measurement at some time $t_0,$ an experiment with measurements at more than one time point provides additional information that may serve to relax the limitation.
For these reasons, it seems that the number of derivatives needed to build $\O_I$ is not necessarily the number of derivatives that need to be estimated for practical purposes.

\begin{figure}[htb]
    \centering
    \includegraphics[width=1.0\textwidth]{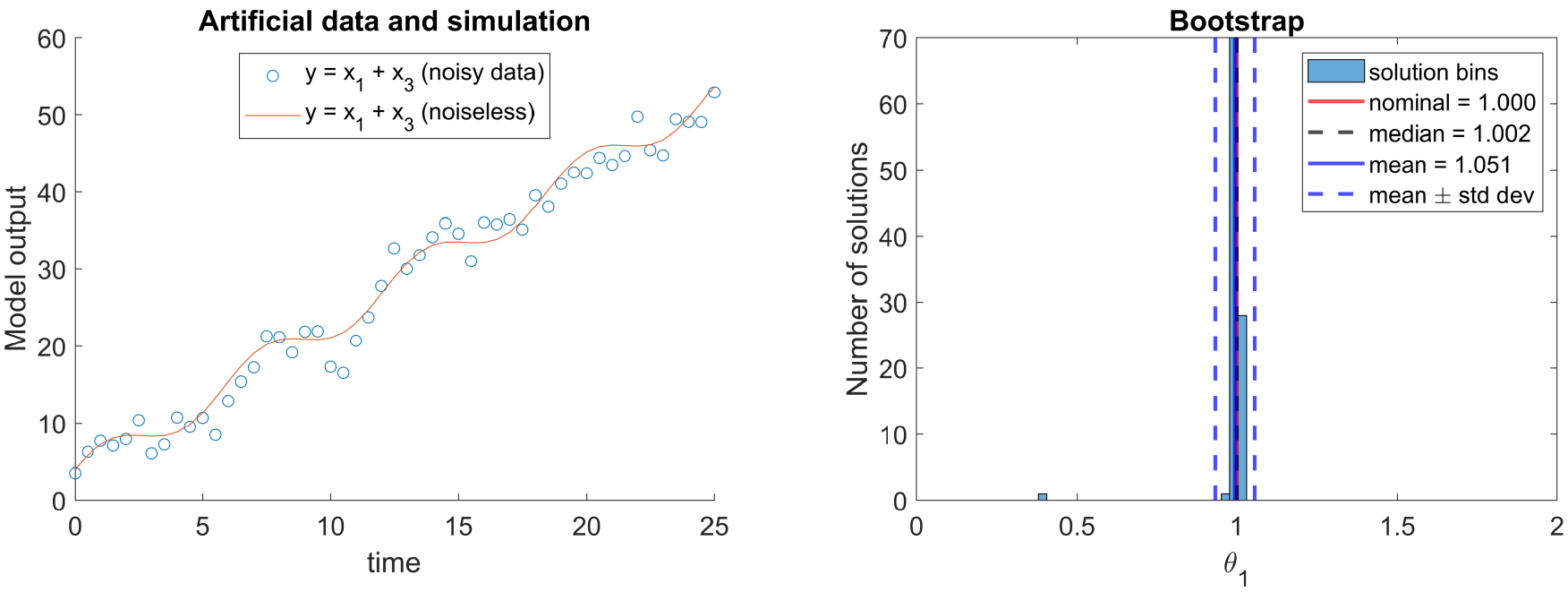}
    \caption{Model $\mathcal{M}_{3.4}$. The panel on the left shows the noisy data (blue circles) of one of the five experiments used for parameter estimation, as well as the noiseless simulation of the output (red line). The panel on the right shows the resulting \revision{bootstrap} of parameter $\theta_1$. It can be seen that the parameter is practically identifiable.}
    \label{fig:ex21}
\end{figure}

To support this claim, we perform the following numerical experiment. We generate synthetic data by simulating model $\mathcal{M}_{3.4}$ from five different experimental conditions, adding \revision{Gaussian noise with zero mean and a standard deviation of 2.0.} This level of noise is one that we could reasonably expect in many practical situations. Figure \ref{fig:ex21} shows in its left side the noiseless signal along with the corresponding noisy pseudo-experimental data of one of the experiments; the other four are similar. 
Then, we try to estimate the values of the parameters and of the initial conditions of each experiment. The right hand side of \revision{Figure} \ref{fig:ex21} shows the \revision{bootstrap histogram obtained for $\theta_1$, which shows that $\theta_1$ is clearly identifiable, with median and mean estimates almost identical to the nominal value ($\theta_1=1$) and a very low standard deviation. 
}

\subsection{\revision{Numerical identifiability and observability through singular value decomposition}}

The rank tests discussed so far provide a binary answer to the observability/identifiability question: either a given variable is observable, or it is not. The same can be said for a model as a whole. However, if the elements of $\O_I$ are not symbolic variables but numerical values, it is possible to provide also a numerical answer, which informs about practical observability. 

The idea is to perform a singular value decomposition (SVD) of the numerical $\O_I$. If the model is practically observable, all the singular values will be of similar magnitude, i.e. the condition number (the ratio of the largest singular value over the smallest one) will be small \cite{dochain1997modelling}. Likewise, the smallest singular value indicates the worst estimate, so ideally it should be large.
The idea of analysing a \textit{linear} system in this way was brought to the attention of the systems and control community by \cite{moore1981principal}.
\revision{\cite{stigter2021computing} used a similar approach, but performing SVD of a sensitivity matrix instead of an observability-identifiability one. 

To illustrate this methodology, we will recall the previously considered examples $\mathcal{M}_{3.2}$ and $\mathcal{M}_{3.4}$, as well as the following model:
\begin{equation}\label{eq:big}
\mathcal{M}_{3.5}:
    \left\{%
    \begin{array}{ll}
	\dot{G}     &= u-(c+s_i\cdot I)\cdot G,\\
	\dot{\beta} &= \beta\cdot \left(\dfrac{1.4583\cdot 10^{-5}}{1+\left( \dfrac{8.4}{G}\right)^{1.7}}-\dfrac{1.7361\cdot 10^{-5}}{1+\left( \dfrac{G}{4.8}\right)^{8.5}}\right),\\
	\dot{I}     &= \beta\cdot \dfrac{G^2}{\alpha^2+G^2} -\gamma\cdot I,\\
	y           &= G.
    \end{array}%
    \right.
\end{equation}

This model represents the regulation of plasma glucose concentration, $G,$ by means of insulin ($I$), which is secreted by $\beta$ cells. This model has been slightly modified from the one presented by \cite{karin2016dynamical}, by removing one parameter ($p$) and transforming one variable ($\beta$), as well as by considering $s_i$ as a known constant. While the original model was structurally unidentifiable, the one given by \eqref{eq:big}, which has three unknown parameters ($c,\alpha,\gamma$), is SLI.

Thus, we have three models ( $\mathcal{M}_{3.2}$,  $\mathcal{M}_{3.4}$, and $\mathcal{M}_{3.5}$) that are structurally locally identifiable and observable. Let us now calculate their observability-identifiability matrices, $\O_I$, and replace the symbolic variables with their nominal numerical values. The SVD of the resulting arrays yields the singular values plotted in logarithmic scale in Figure \ref{fig:svds}.
None of these plots exhibits a very large gap between singular values, and the largest condition number ($2.2\cdot 10^6$) is that of $\mathcal{M}_{3.2}$. However, when we compare their absolute values, we find that the smallest ones are in the order of $10^{-1}$ for the first two examples, while for $\mathcal{M}_{3.5}$ they decrease to $10^{-4}$ and $10^{-5}$. These values suggest that  $\mathcal{M}_{3.5}$ has practical identifiability issues. 

\begin{figure}[htb]
    \centering
    \includegraphics[width=1.0\textwidth]{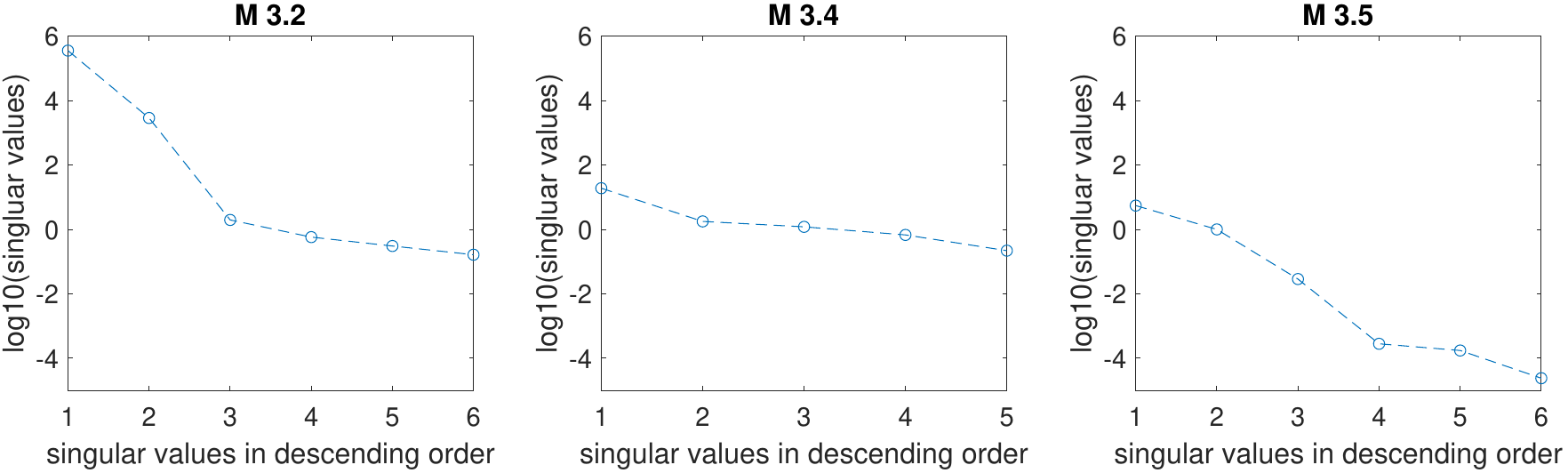}
    \caption{Singular values of models $\mathcal{M}_{3.2}$,  $\mathcal{M}_{3.4}$, and $\mathcal{M}_{3.5}$, in logarithmic scale.
    }
    \label{fig:svds}
\end{figure}

To test this prediction, we explore the practical identifiability of this model by performing a bootstrap procedure, as in the previous cases. Simulating the model with nominal values, we generate 100 pseudo-experimental realizations adding Gaussian noise with zero mean and 0.2 standard deviation, and we solve the 100 optimization problems. Figure \ref{fig:big_sim} shows the fit obtained for one realization; the others were similar. Figure \ref{fig:big_boots} plots the histograms of bootstrap solutions for all the estimated variables: three unknown parameters, $c, \alpha, \gamma,$ and the three initial conditions, $G(t=0), I(t=0), \beta(t=0)$. The values of two of the parameters, $c$ and $\alpha,$ and of the initial condition of the measured variable, $G,$ are well estimated. In contrast, parameter $\gamma$ is clearly practically unidentifiable, and the estimates of the remaining initial conditions are highly uncertain too. Thus, the SVD of $\O_I$ correctly reports the existence of practical identifiability issues.
The SVD also gives correct results for the previous two case studies, which we found were structurally and practically identifiable. 

\begin{figure}[htb]
    \centering
    \includegraphics[width=1.0\textwidth]{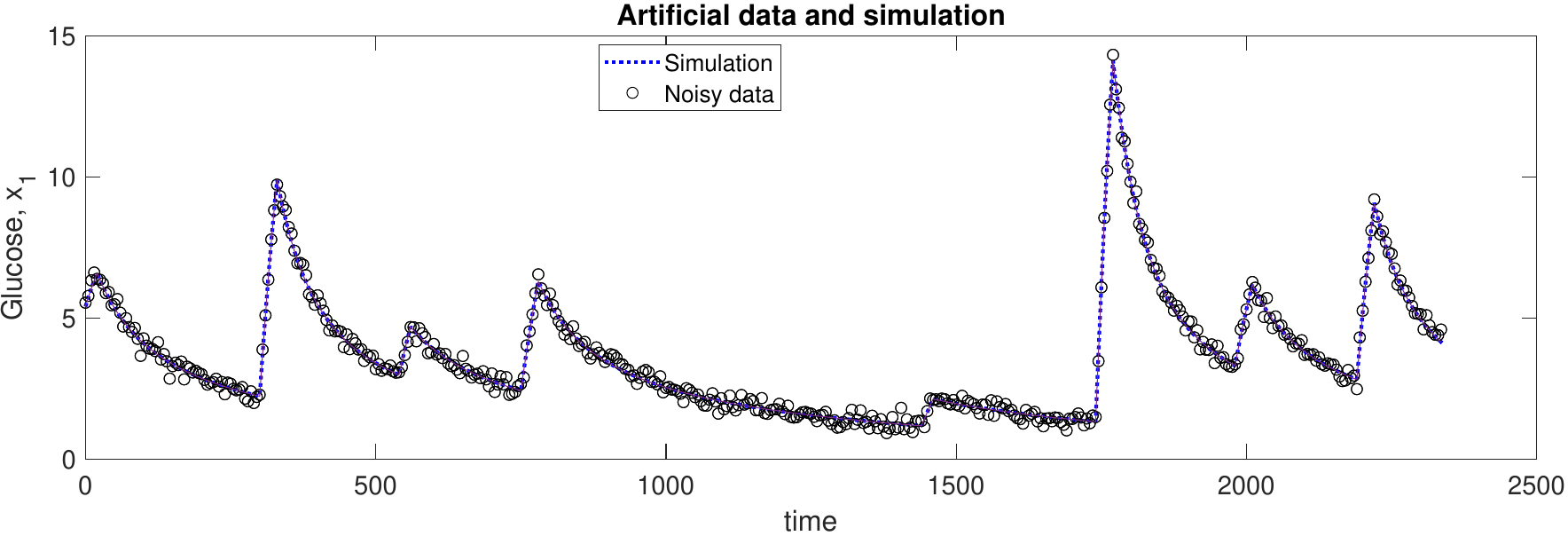}
    \caption{\revision{Model $\mathcal{M}_{3.5}$. Noisy data (circles) of the experiment used for parameter estimation, as well as the noiseless simulation of the output, glucose.}
    }
    \label{fig:big_sim}
\end{figure}

\begin{figure}[htb]
    \centering
    \includegraphics[width=1.0\textwidth]{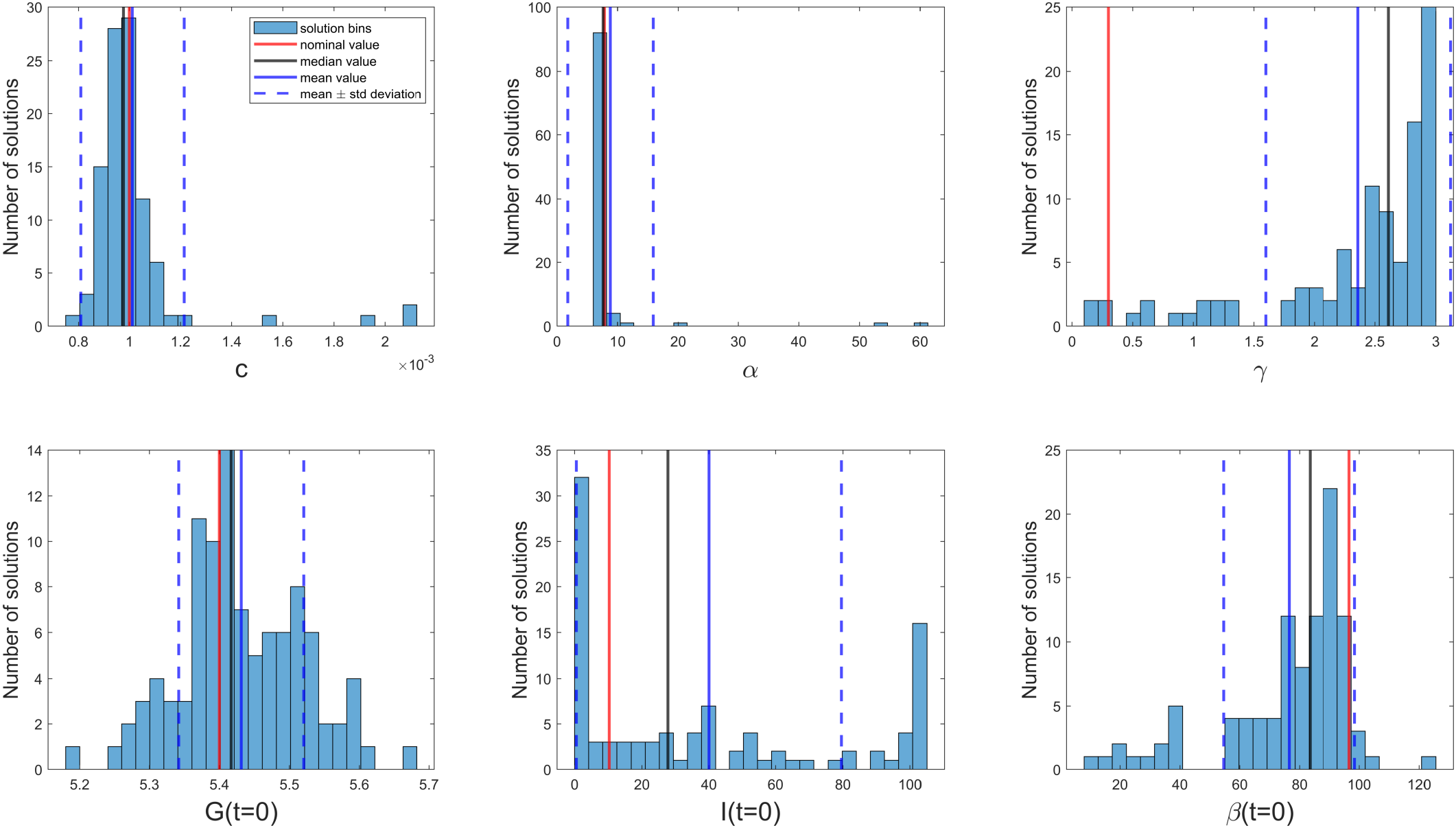}
    \caption{\revision{Model $\mathcal{M}_{3.5}$. Bootstrap results: histograms of all estimated variables.}
    }
    \label{fig:big_boots}
\end{figure}

}

\section{Conclusions}\label{sec:conclusions}

Th observability rank condition (ORC) is one of the classic approaches for the study of structural local identifiability and observability. It is based on the construction of a symbolic observability-identifiability matrix using a differential geometry concept, the Lie derivative. The rank of this matrix provides a binary assessment of (structural, local, single experiment) identifiability and observability. In its common form, the ORC does not take into account certain aspects that affect these properties from a practical point of view. In this paper we have seen how the ORC can be modified in order to take some of these factors into account. 

Specifically, (i) the additional information provided by multiple experiments can be incorporated by creating replicates of certain variables, and the resulting test correctly informs about improvements in identifiability. (ii) Certain characteristics of the inputs, such as their time-varying nature, can be specified by setting some of their time derivatives to zero. (iii) A more nuanced measure of observability can be obtained by performing singular value decomposition on a numerical specialisation of the observability-identifiability matrix. 
In regard to aspects (i) and (ii), we have seen that -- at least in some cases -- piecewise constant inputs are better approximated by ramps than by multiple experiments with constant inputs, since in the latter description some information about the transitions is lost.

We have also discussed several cases in which similar extensions, albeit intuitively reasonable, yield misleading results: (iv) building the observability-identifiability matrix with a certain number of derivatives is not equivalent to assuming that we must be able to estimate that same number of derivatives of the output from data; in practice, the fact that multiple experiments and/or multiple time point measurements provide additional information can relax said assumption. (v) When the ORC is evaluated for numerical specialisations of the initial conditions, it may be necessary to compute additional Lie derivatives to achieve full rank, and therefore rank deficiency does not necessarily imply unobservability; in this case the test provides a sufficient but not necessary condition.

To the best of our knowledge, the way in which issues (iv) and (v) can be addressed in the framework studied in this paper remains open. 
Likewise, future work could address the incorporation of other practical aspects, such as e.g data noise.

\section*{Acknowledgement}

I thank the mathematical research institute MATRIX in Australia where part of this research was performed.
I also acknowledge funding from 
grant PID2023-146275NB-C21 funded by MICIU/AEI/10.13039/501100011033 and ERDF/EU,
and grant RYC-2019-027537-I funded by MCIN/AEI/ 10.13039/501100011033 and by ``ESF Investing in your future''. 


\end{document}